\documentclass[12pt]{extarticle}                                          
\usepackage[T1]{fontenc}                                                       
\usepackage[utf8]{inputenc}                                
\usepackage[english]{babel}                                                    
\usepackage{fontenc}                           
\usepackage[letterpaper]{geometry}                                              
\usepackage{amsfonts,amssymb,amsmath}                                           
\usepackage{graphicx}                                                           
\usepackage{animate}                                                            
\usepackage[svgnames]{xcolor}                                                   
\usepackage{enumitem}                                                      
\usepackage[colorlinks=True,linkcolor=blue,citecolor=blue,urlcolor=blue]{hyperref}                                                           
\usepackage[squaren,Gray]{SIunits}
\usepackage{geometry}
\usepackage{etoolbox}

\usepackage{array,multirow,makecell}
\usepackage{tabularx}           
\usepackage{subcaption}

\usepackage[sort]{natbib}
\makeatletter
\newcommand\footnoteref[1]{\protected@xdef\@thefnmark{\ref{#1}}\@footnotemark}
\makeatother

\begin{document}
%
%\begin{titlepage}
%\centering
%{\scshape Intersnship Report - Master Astronomie Astrophysique et  Ingénierie Spatiale \par}
%\vspace{2cm}
%{\Huge \bfseries Radio and UV emission of cocoon in Gamma Ray Bursts\par}
%
%%\includegraphics[width=0.7\textwidth]{orion.jpg}\par\vspace{1cm}
%
%\vspace{2cm}
%
%\begin{tabular}{rcl}
% \Large Odélia Teboul & \hskip 2cm  &  \Large Tsvi Piran \\ 
% \large odelia.teboul@obspm.fr  &  & \large tsvi.piran@mail.huji.ac.il 
%\end{tabular}
%
%
%\vspace{0.5cm}
%%\includegraphics[width=0.7\textwidth]{orion.jpg}\par\vspace{1cm}
%\end{titlepage}
\begin{titlepage}
\vspace*{-3cm}
\centering
{\scshape Intersnship Report - Master Astronomie Astrophysique et  Ingénierie Spatiale \par}
\vspace{2cm}
{\Huge \bfseries Emission of cocoon afterglow for short Gamma Ray Bursts  \par}
\vspace{1cm}

{\Large \hspace{-1.5cm} Odélia Teboul \hspace{3.5cm}  Tsvi Piran\par}
{ \Large odelia.teboul@obspm.fr  \hspace{2cm} tsvi.piran@mail.huji.ac.il \par}
\vspace{1.5cm}
\includegraphics[scale=0.1]{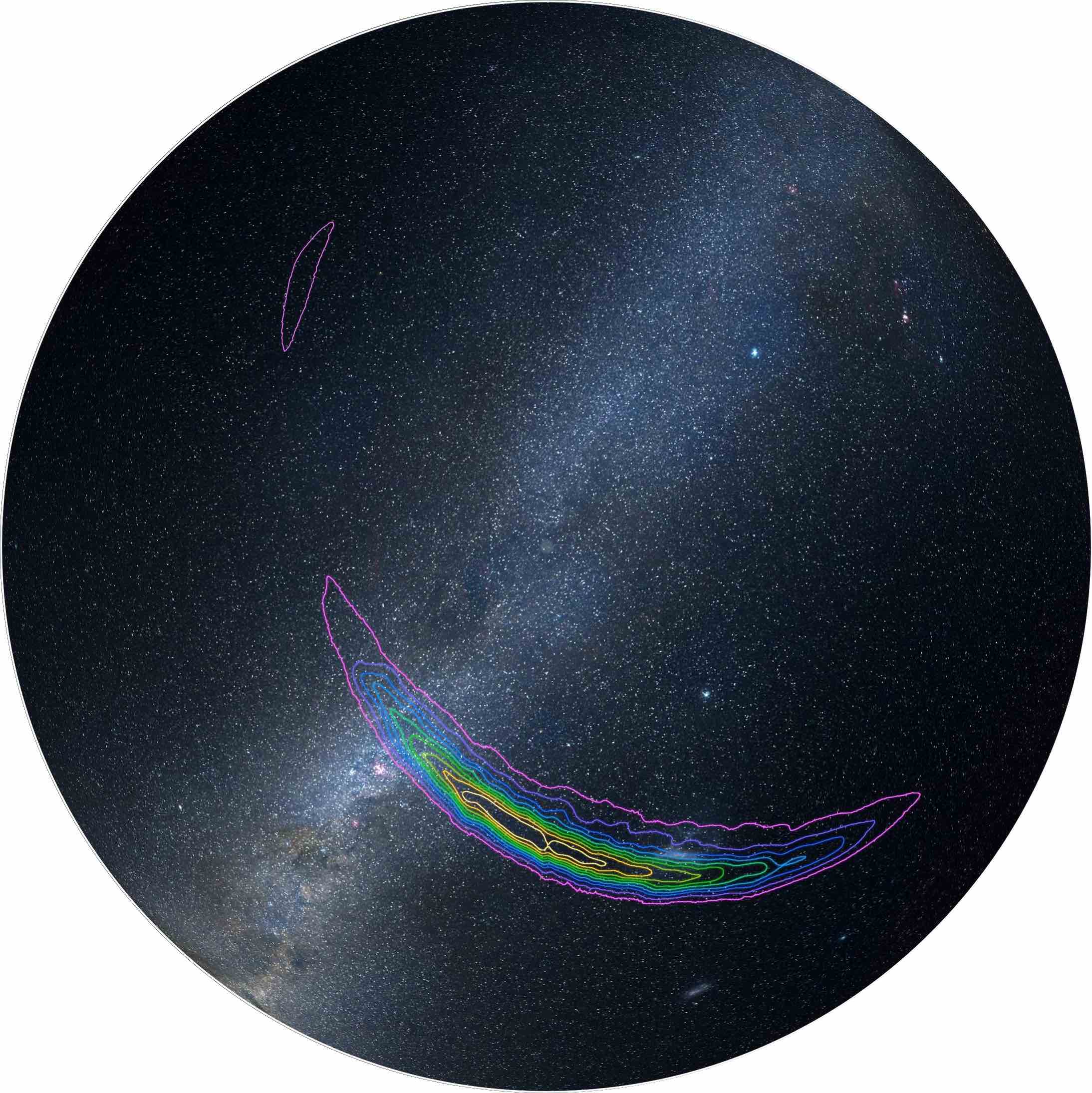}

{ \Large  23 Juin 2017\par}
\vspace{1cm}
\begin{figure}[h]
\begin{subfigure}{.5\textwidth}
\raggedright
  \includegraphics[scale=0.12]{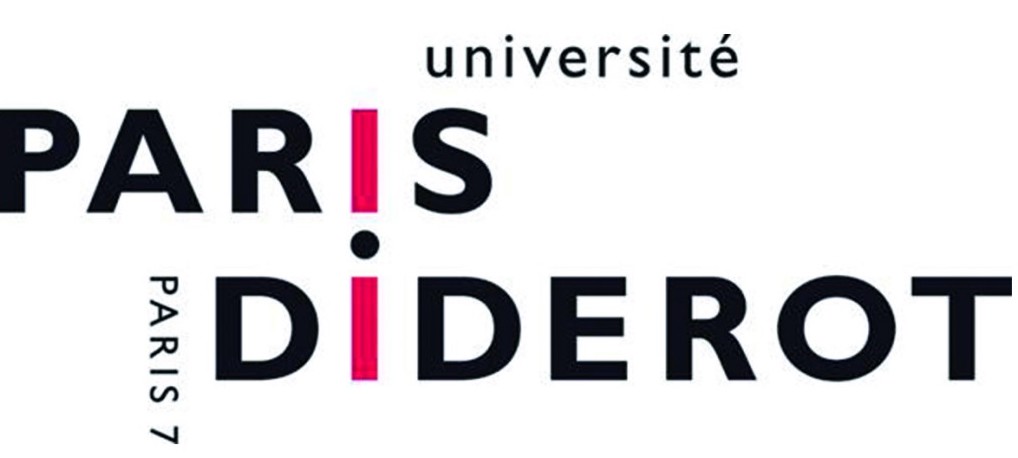}
\end{subfigure}
  \begin{subfigure}{.5\textwidth}
\raggedleft
  \includegraphics[scale=1.0]{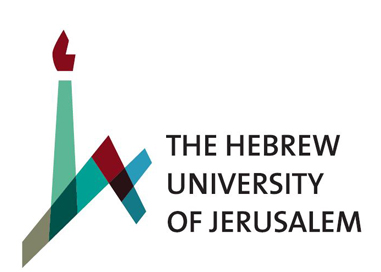}
\end{subfigure}
\end{figure}
\vspace*{-3cm}
\end{titlepage}

\begin{titlepage}
{\bf Abstract} \\ \\
The three gravitational wave events detected by LIGO are opening a new era for high-energy astrophysics. Nevertheless, location of such events remain unknown. A promising solution to the localization problem is to find an electromagnetic (EM) counterpart of GW-generating events, such as binary neutron star mergers (BNS). Indeed, their GW emission will be above sensitivity threshold in the near future. BNS are also considered as short Gamma Ray Bursts (sGRB) progenitors. However, sGRB are highly beamed. In this study, we will therefore focus on another EM counterpart candidate: the cocoon afterglow. The propagation of the GRB jet inside the matter ejected by the BNS produces a cocoon. Then, similarly to the GRB afterglow, a cocoon afterglow is produced, but with a mildly relativistic velocity. Firstly, we propose a model that gives the full hydrodynamic evolution of the cocoon including the mildly relativistic regime. Then we calculate the cocoon afterglow emission in X-ray, optical and radio wavelengths. Finally, we compare the cocoon afterglow emission to the GRB afterglow emission.
\end{titlepage}
%\addtocounter{chapter}{1}
%\newpage
%\tableofcontents
%\newpage
\begin{titlepage}
{\bf Acknowledgments} \\ \\

I would like to thank Tsvi Piran for these very interesting three months and it will be a pleasure to work together during my PhD.

I would also like to thank Nir Shaviv for his wonderful help. I am glad that we will work together during my PhD.

I would also add special thanks to Noemie Globus for her advice and help.

Finally, I would like to thank all the students and professors of the Astrophysics and Cosmology team of the Racah for this great experience. 
\end{titlepage}

\tableofcontents

\newpage

\section{Introduction}
\label{Introduction}

%\subsection{Motivation}
%In 1967, VELA detect an intense burst in gamma rays : gamma rays bursts 
%Since their discovery in 1967 by VELA satellite, Gamma Ray Burst (GRB) have been one of the « tigger » point for high energy astrophysicist. 
%separation short and long 
%The afterglow,  discovered in 1997 by Beppo-SAX a brader information. Since Swift the puzzling

\subsection{Motivation : Gravitational waves localization  }
\label{Intro-Motivation}
\begin{figure} [h!]
  \centering
  \includegraphics[scale=0.4]{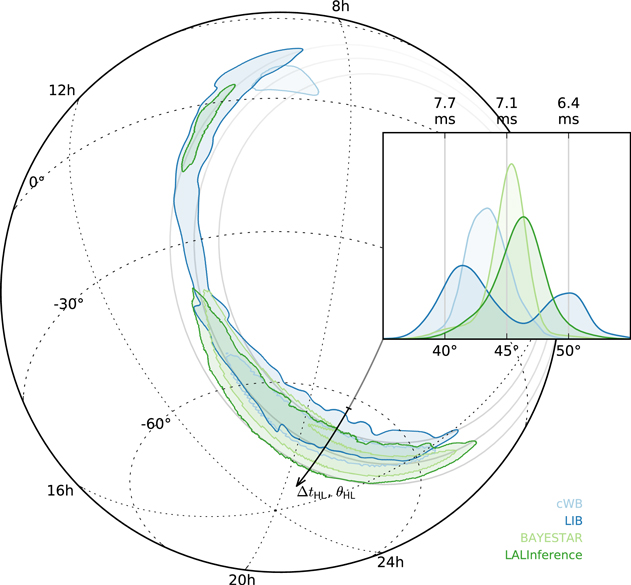}
  \caption{Locatization error of GW151226 (Credit : \citealt{LIGO-FirstDetection})}
  \label{fig:GW} 
\end{figure}

The recent discovery of  gravitational waves (GW) is opening up new horizons for high-energy astrophysics.  Since GW interact very weakly with matter, unlike electromagnetic radiation, they can carry information from the early Universe and its origin as well as other unseen high energy phenomena.  Moreover, GWs can provide new tests of general relativity, especially in the dynamically strong-field regime, as is the case with the three recent detections of Binary Black-Hole merger events by the Laser Interferometer Gravitational Wave Observatory (LIGO) experiment. However, the location of such events remains undetermined, as experimental triangulation of GW is currently impossible (see fig.\ \ref{fig:GW}). One promising lead is to look for electromagnetic (EM) counterparts. Finding an EM counterpart of GW will allow the localization of  the events.
To date, the three events detected by LIGO are binary black hole mergers \citep{LIGO-FirstDetection,Abbott2017}. Nevertheless, GW emitted during binary neutron star mergers will be above the sensitivity threshold of Advanced LIGO for the next campaigns.

Binary neutron star mergers have also been recognized as the possible progenitors of short gamma-ray bursts (GRBs) \citep{Eichler1989,Nakar2007}. If so, short GRBs (sGRB) and their afterglows could give rise to fascinating electromagnetic counterparts to GW. 
However, it is not clear to what extent can either the GRB or the GRB jet afterglow be observable given beaming. Hence, we have to look for other EM counterparts. 
So far, macronova which behave like r-process supernova and radio flares are widely studied in the literature \citep{Metzger2017}. Here we discuss an additional EM counterpart that arises from the interaction of a cocoon formed during the jet propagation within matter surrounding the merger.

\subsection{Cocoon}
\label{Intro-cocoon}
When two neutron stars merge, matter is ejected prior to the jet’s onset by winds driven from the newly formed hyper-massive neutron star and from the debris disk that forms around it \citep{Hotokezaka2015}.
  
\begin{figure}[h!]
  \centering
  \includegraphics[scale=0.5]{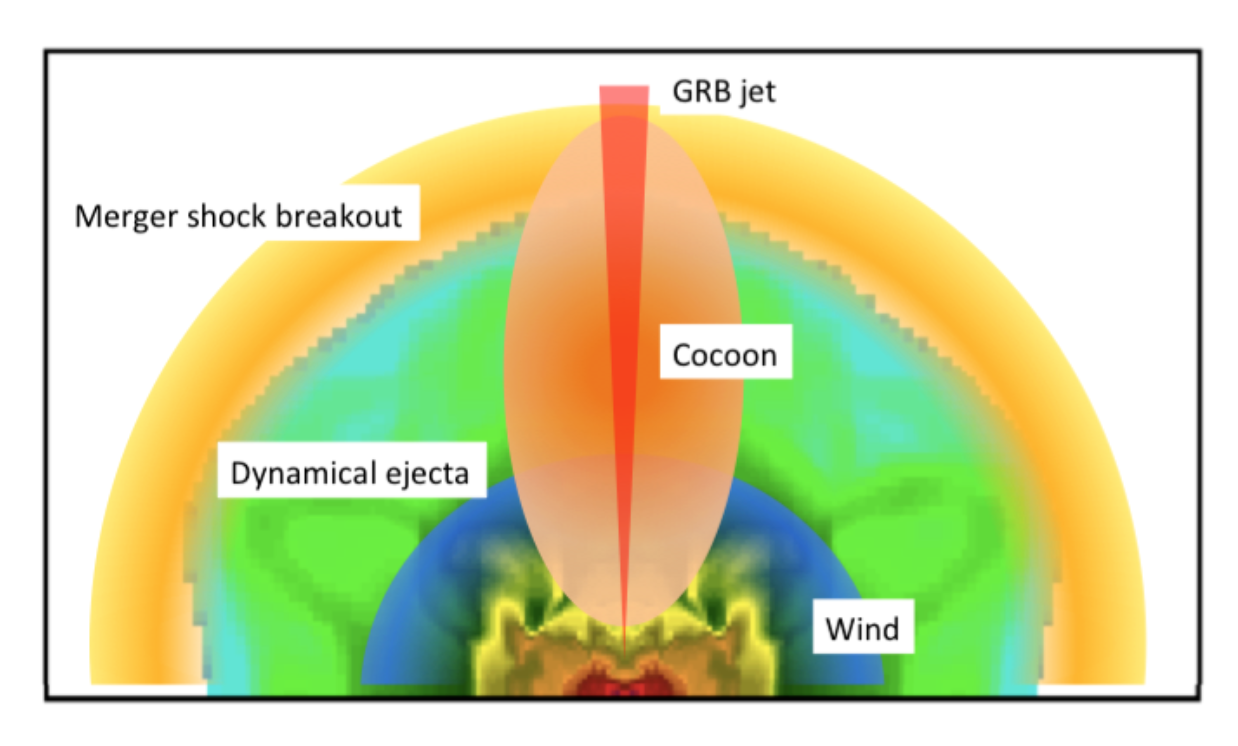}
  \caption{Heuristic description of the jet, cocoon, wind and dynamical ejecta following a NS merger, presumably producing a sGRB.}
  \label{fig:cocoon}
\end{figure}

The cocoons are generated during the interaction of the GRB jet with this surrounding matter \citep{Nagakura2014,MB2014}.
The different components of the ejecta for sGRB are presented in fig.\ \ref{fig:cocoon}. 
\\The energy of the cocoon is comparable to the energy of GRB which includes the prompt emission and afterglow kinetic energy \citep{Nakar2017}. In addition, the cocoon afterglow will behave like the GRB afterglow \citep{Piran2004}.
As shown in this figure, the cocoon opening angle is wider than the jet opening angle. Hence, when observing off jet-axis, sGRB cocoon signatures could be of prime importance. 
Nevertheless, the initial Lorentz factor of the cocoon  is  $\gamma_{0} \approx  10 $. Therefore, when expanding, the cocoon will rapidly reach the mildly relativistic regime for which neither the Sedov-Taylor \citep[ST][]{Sedov1958,Taylor} solution nor its fully relativistic counterpart by  Blandford and McKee \citep[BM][]{BM} are well applicable. We propose here a model that allows a smooth transition from the BM phase to the ST phase.
 
\newpage
\section{Cocoon}
 \label{Cocoon}
Hereafter we discuss, the cocoon formation demonstrated in both analytical and simulation works. We also present the cocoon properties.
  
\subsection{Cocoon formation : analytical and simulation results }
 \label{Cocoon-formation}
\begin{figure} [h!]
  \centering
  \includegraphics[scale=0.5]{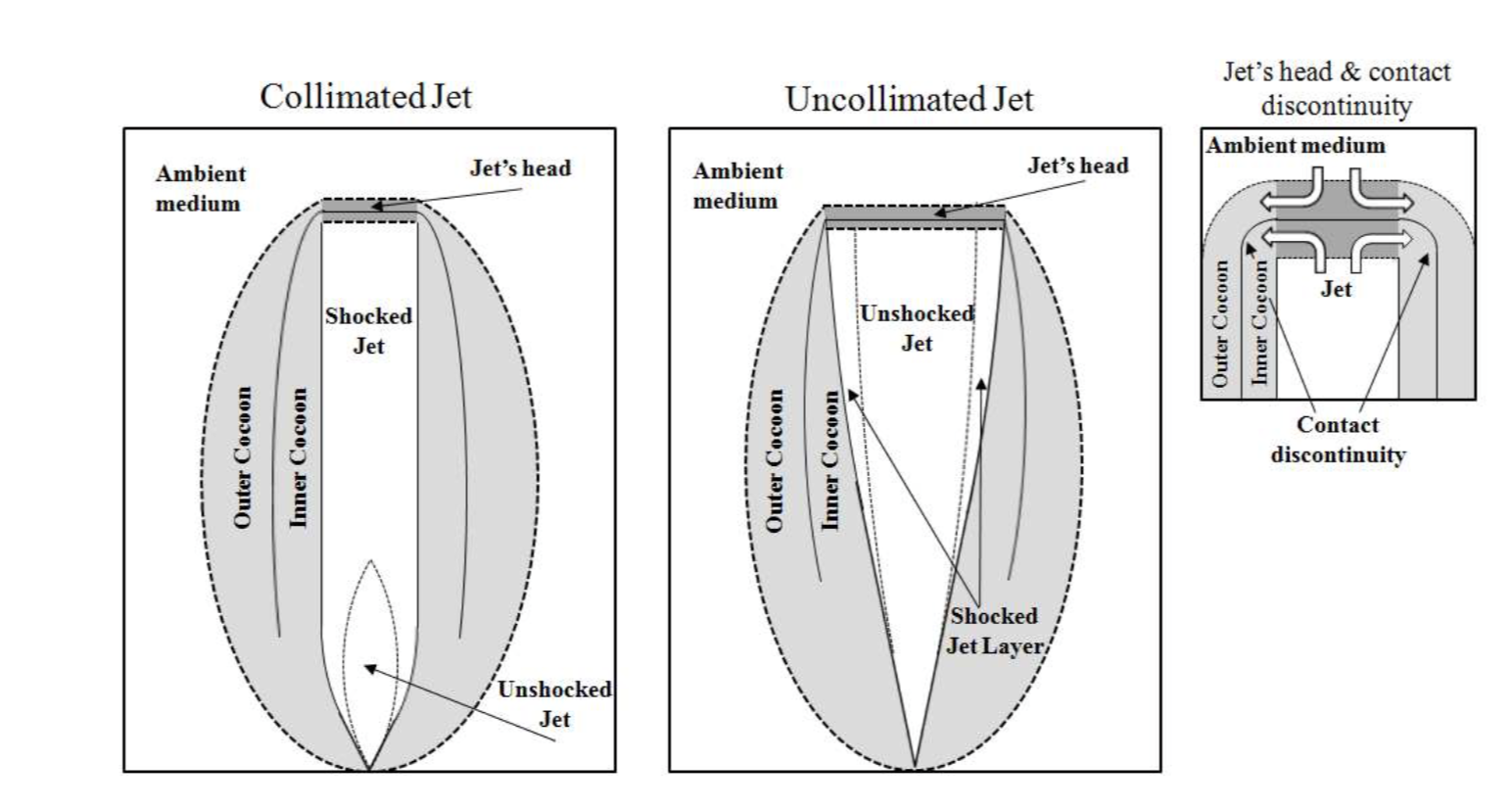}
  \caption{Cocoon formation,  Credit \cite{Bromberg2011}}
  \label{fig:CF}
\end{figure}

 It has been shown, for active galactic nuclei, that the propagation of the jets they accelerate through surrounding media generates a double bow-shock structure at the head of the jet  \citep{Blandford1974,Scheuer1974}. Energy and matter that enter this structure are pushed aside due to a high pressure gradient and create a hot cocoon around the jet. The cocoon, in turn, applies pressure on the jet and compresses it. 
The cocoon formation was studied analytically by \citep{Bromberg2011}, and it is described heuristically in fig.\ \ref{fig:CF}. 

\begin{figure} [h!]
  \centering
  \includegraphics[scale=0.4]{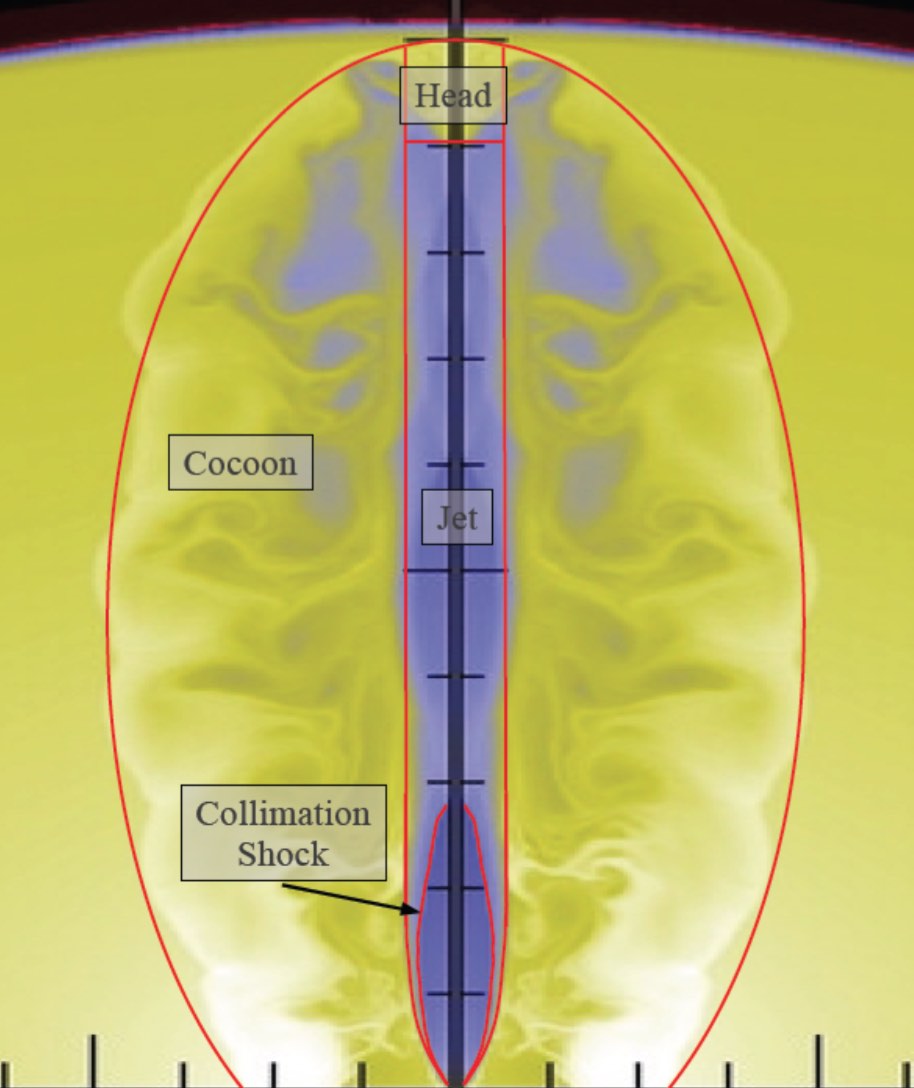}
    \caption{Cocoon Simulation,  Credit \cite{Mizuta2009}}
  \label{fig:CS}
\end{figure}

Meanwhile, simulations of GRB jet propagation inside a star were conducted as well (e.g. \citealt{Zhang2003}, \citealt{Morsony2007}, \citealt{Mizuta2009}). All these analyses demonstrated the formation of a jet head and a hot cocoon. See for instance the simulations by \cite{Mizuta2009}, also depicted in fig.\ \ref{fig:CS}.
Therefore, we can conclude that for both short and long GRBs, the interaction of the jet with the  surrounding matter generates a cocoon. Long GRBs arise at the end of massive star life after the collapse of the core (citer collapsar model). In this case, a cocoon would be generated by the interaction of the jet with the progenitor star before the breakout \citep{Nakar2017}.
Note that the following study focuses on the cocoon in the case of short GRBs, but could also be applied for long GRB by appropriately considering different energy and initial Lorentz factor.

\subsection{Properties of the cocoon}
 \label{Cocoon-Properties}
\subsubsection{Energy}
 \label{Cocoon-Properties-Energy}
The total cocoon energy, $E_{0}$, which is the total energy deposited by the jet in the cocoon until the breakout time $t_{b}$, is expected to be comparable to the total GRB energy given by the sum of the prompt emission and afterglow kinetic energy. The reason is that the typical breakout time is comparable to the typical burst duration (Bromberg et al. 2012) and the jet deposits almost all its energy into the cocoon during its propagation in the ejecta. The total  cocoon energy  $E_{0}= \int_{t_{inj}}^{t_{b}}L_{j}(1-\beta _{h})dt$ where $L_{j}$  is the total two sided luminosity  and $\beta _{h}c$ is the velocity of the jet's head. 
Note that while the jet is relativisitc the jet's head velocity is typically of order of $0.1-0.3c$ \citep{Matzner2003,Bromberg2011}.  The GRB’s energy is the jet’s energy after the breakout and there is no reason to expect that the jet will not have the same luminosity before and after breakout. It has been shown by \cite{Moharana2017} that the distribution of sGRB durations suggests that the jet is launched for at least a few hundred milliseconds in order for it to break out of the ejecta. Therefore, we can approximate $E_{0}\sim L_{j}(t_{b} - t_{inj} - R_{bo}/c) $ where $R_{bo}$ is the radius at time of the breakout. 

In such a case the cocoon carries an energy that is comparable to that of the sGRB itself and the cocoon breakout radius is around $ 10^{9}$~cm.
\subsubsection{Composition and Lorentz factor}
 \label{Cocoon-Properties-Composition}
As shown in section \ref{Cocoon-formation},  the cocoon is composed of an inner and an outer part. The inner part is made of jet material while the outer part is made of ejected matter from the binary neutron star merger. Nevertheless, these two parts can be partially to fully mixed and the mixing ratio will change the Lorentz factor of the cocoon \citep{Nakar2017}. 
Here we focus on the inner part of the cocoon which is made of jet material and hence does not contain too much mass. This component will be relativistic, we consider here $\gamma_{0} \approx  10$. Therefore this inner part is more likely to create an afterglow that will behave like a GRB afterglow \citep{Piran2004}. 

\section{Hydrodynamics of the shock}
 \label{Hydrodynamics}
Similarly to GRB afterglows, a cocoon afterglow arises from the interaction of the cocoon with the matter surrounding it. This interaction is mainly hydrodynamical.

\subsection{Evolution of Lorentz factor of the cocoon}
 \label{Hydrodynamics-Evolution}
\begin{figure} [h!]
  \centering
  \includegraphics[scale=0.7]{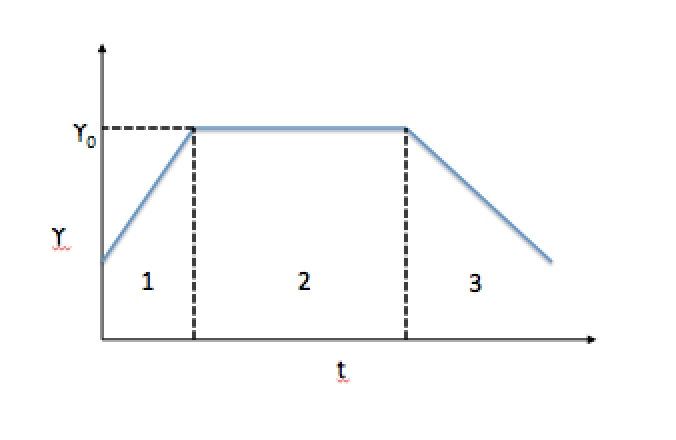}
  \caption{Evolution of Lorentz factor of the cocoon}
  \label{fig:Lorentz}
\end{figure}

The evolution of the cocoon can be separated into 3 distinct phases shown in. fig.\ \ref{fig:Lorentz}, which represent 3 steps of the blast wave evolution, as described below. 
\\The first phase can be described by the fireball model, proposed by \cite{Paczynski1986} and \cite{Goodman1986}. They have shown that the sudden release of a large quantity of gamma ray photons into a compact region can lead to an opaque photon–lepton “fireball” through the production of electron–positron pairs. The term “fireball” refers here to an opaque radiation–plasma whose initial energy is significantly greater than its rest mass. Goodman  considered the sudden release of a large amount of energy, $E_{0}$, in a small volume, characterized by a radius, $R_{0}$ which could  occur in an explosion.
They showed that if the ejecta stays optically thick long enough, then all the internal energy can be converted into kinetic energy, allowing the matter to reach a final Lorentz factor $\gamma_{0}$. Here we consider for the cocoon, $\gamma_{0} \approx  10$.

Subsequently, during the second phase, the fireball expands and collects an exterior mass $M_{ext}$ until the exterior mass $M_{ext}\approx {M_{0} / \gamma_{0}}$.  This occurs at a radius \citep{Daigne}:
\begin{equation}
R_{dec}\approx \left (\frac{3E_{0}}{4\pi \gamma _{0}^{2}n_{0}m_{p}c^{2}}  \right )^{1/3}.
\end{equation}
Considering an initial energy  $E_{0} = 10^{49}$ erg, $\gamma_{0} \approx 10 $ and $n_{0} = 1\: $   cm$^{-3}$   we obtain $R_{dec}\approx 2.5 \: 10^{16}$cm.

Finally, during the third phase, after  reaching $R_{dec}$ the blast waves decreases following Blandford and McKee and later on Sedov and Taylor. This part will be discussed into details in  section.

\subsection{Shock properties}
\label{Hydrodynamics-Shock}
\subsubsection{Shock : general properties}
\label{Hydrodynamics-Shock-general}
Consider the situation when a cold relativistic shell (whose internal energy is negligible compared to its rest mass) moves into the cold interstellar medium (ISM). Conservation of mass, energy and momentum determine the Hugoniot shock jump conditions across the relativistic shocks for the case when the upstream matter is cold (see e.g. \citealt{BM}).
 \begin{figure} [h!]
  \centering
  \includegraphics[scale=0.7]{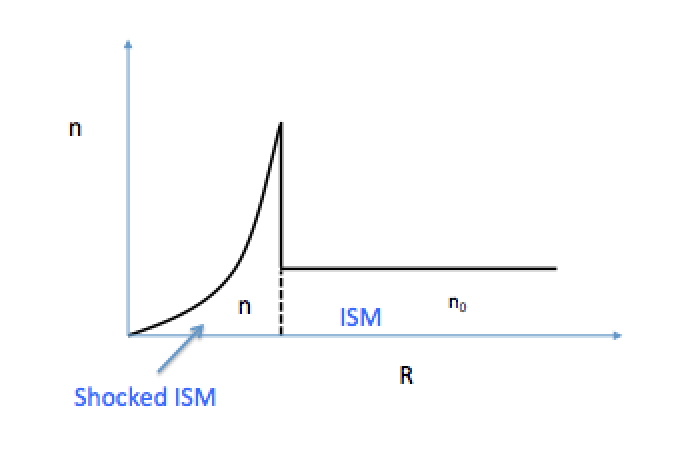}
  \caption{Shock characteristics}
  \label{fig:shock}
\end{figure}
\\Note that the above figure presents the shock for a given $\gamma$, and that as described below, $\gamma$ varies with time generating different shocks with different characteristics.

For the shocked ISM the particle density is $n$  while  $n_{0}$ is the exterior density. 
With $\gamma$ the Lorentz factor of the shocked fluid, the particle density, energy   of the shocked ISM are defined  following Blandford and McKee 1976 as : 
\begin{eqnarray}
n &=& 4 n_{0} \gamma,  \\
E &=&  N m_{p} c^{2} (\gamma -1),
\end{eqnarray}
with  $N = n\frac{4\pi }{3}R^{3}$, number of particles of the shocked ISM.
 
\subsubsection{Shock : acceleration of the electrons}
\label{Hydrodynamics-Shock-acceleration}
We assume that electrons are accelerated in the shock to a power law distribution of Lorentz factor $\gamma_e$, with a minimum Lorentz factor $\gamma _{m}$ such as $ N(\gamma _{e}) d\gamma _{e}  \alpha  \gamma _{e}^{-p} d\gamma _{e},  \gamma _{e} \geq  \gamma _{m}$,
with $p>2$, in order to keep  the energy of the electrons finite.
\\
\\
We consider $ p=2.5$ in relativistic regime and $ p=3$ in non relativistic regime \citep{Sari1996}. In our model during the transition phase $p$ evolves linearly between $p = 2.5$ and $p = 3$. 
\\
\\
Let $\varepsilon _{e}$ be the fraction of the shock energy going into the electron energy density: 
\begin{equation}
 E_{e} = \varepsilon _{e} E, 
 \end{equation} 
with $ E_{e}$ energy density of the electrons. 
\\
\\
Considering that a constant fraction $\varepsilon _{e}$ of the shock energy  goes into the electrons, we get  for the relativistic regime \citep{Sari1998}: 
\begin{equation}
\gamma_{m}=\epsilon _{e}\frac{p-2}{p-1} \frac{m_{p}}{m_{e}}\gamma
\end{equation} 
where $m_{p}$ is the mass of proton and $m_{e}$ the mass of the electron.

However, during the non relativistic regime, assuming that the same constant constant fraction $\varepsilon _{e}$ of the shock energy  goes into the electrons, we obtain : 
\begin{equation}
\gamma_{m}=\epsilon _{e}\frac{p-2}{p-1} \frac{m_{p}}{m_{e}}\beta^{2} 
\end{equation}
\subsubsection{Magnetic field of the shocked ISM}
\label{Hydrodynamics-Shock-magneticfield}
Similarly, let $\varepsilon _{B}$ be the fraction of the shock energy going into magnetic energy density:
\begin{equation}
 \frac{B^{2}}{8\pi} = \varepsilon _{B} E.
 \end{equation}
We assume that a constant fraction $\varepsilon _{B}$ of the shock energy goes into magnetic energy density. Therefore the magnetic field strength is given by \citep{Sari1998}:
\begin{equation}
B = (32\pi m_{p}\varepsilon _{B}n)^{1/2}\gamma c
\end{equation}

\subsection{Deceleration of the blast wave}
\label{Hydrodynamics-Deceleration}

Here, we assume $\varepsilon _{B}=\varepsilon _{e}=0.1$  and consider an  adiabatic evolution. 
%Indeed, the radiative solution assumes that all the internal energy created in the shock is radiated. This requires that the fraction of the energy going into the electrons must be large, i.e.  $\varepsilon _{e} \rightarrow1$, which is not the case here.

\subsubsection{Blandford and McKee, Sedov and Taylor}
\label{Hydrodynamics-Deceleration-Blandford}
We consider a spherical blast wave of radius $R(t)$ propagating into a constant surrounding density $n_{0}$. After $R_{dec}$, the deceleration of the blast wave begins (see fig.\ \ref{fig:Lorentz}), and the evolution of the radius $R$ and the Lorentz factor $\gamma $ while still in the ultra relativistic domain is given by \citep{BM}:
\begin{eqnarray}
\gamma(t) &=& \frac{1}{4}\left ( \frac{17E}{\pi n_{0} m_{p}c^{5}t^{3}} \right )^{1/8} \\ 
R (t) &=& \left ( \frac{17Et}{4 \pi n_{0} m_{p}c } \right )^{1/4}
\end{eqnarray}

When the  non relativistic regime is reached, the evolution of the radius $R(t)$ and the velocity $V (t)$ is then given by \cite{Sedov1958} and \cite{Taylor}.

\begin{equation}
V (t) = \frac{2}{5}\left ( \frac{25E}{4 \pi n_{0} m_{p} } \right )^{1/5} t^{-3/5}
\end{equation}
\begin{equation}
R (t) = \frac{2}{5}\left ( \frac{25E}{4 \pi n_{0} m_{p} } \right )^{1/5} t^{2/5}
\end{equation}

\subsubsection{Transition region}
\label{Hydrodynamics-Deceleration-Transition}
As shown in section \ref{Cocoon-Properties-Composition}, we consider an initial Lorentz factor  $\gamma_{0} \approx  10 $ for the cocoon, therefore, the mildly relativistic regime will be of prime importance. For GRB afterglows, the initial Lorentz factor is $\gamma_{0} \approx  100$, which means that most of the emission will take place during the Blandford and McKee evolution of the blast wave.
However, in our case, we have to determine the evolution of the Lorentz factor during this mildly relativistic to non relativistic transition in which most of the emission of the cocoon will take place. The emission process will be discussed in detail in section \ref{Synchrotron} below. 

For our model, we consider that we have a Blandford and McKee evolution until $\gamma (t) \approx  3$, where we enter the transition region.
In the transition region, the evolution is given by the following extrapolation:
\begin{equation}
S_{trans} = \left ( S_{BM}^{2}  + S_{ST} ^{2}\right )^{1/2},
\end{equation}
where $S_{BM}$ is the BM solution, $S_{ST}$ the ST solution and $S_{trans}$ our solution for the transition regime. Our model allows us to cover the full hydrodynamic evolution of the blast wave by computing a smooth transition between the BM and ST regimes. At the end of the transition region, we have reached a non relativistic regime and therefore the blast wave follows a Sedov and Taylor evolution. Our model will be compared to simulations.

It has been shown \citep{Daigne} that the radius $R_{New}$ at which the blast wave begins to follow the Sedov and Taylor evolution is given by:
\begin{equation}
R_{New}\approx \left (\frac{3E_{0}}{4\pi n_{0}m_{p}c^{2}}  \right )^{1/3}.
\end{equation}
For the given energy and particle density, we get $R_{New}\approx 1.7 \: 10^{17}$~cm.

\begin{figure} [h!]
  \centering
  \includegraphics[scale=0.6]{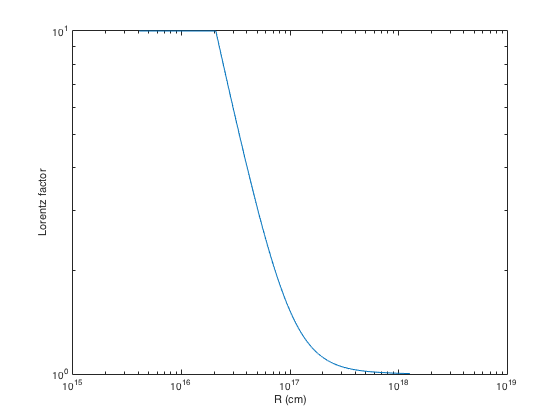}
   \caption{Evolution of Lorentz factor with respect to radius R.}
  \label{fig:gammaR}
\end{figure}

These theoretical values of both $R_{dec}$ and $R_{New}$ are very close to the values obtained with our model, see fig.\ \ref{fig:gammaR}.

\subsection{Beaming}
\label{Hydrodynamics-Beaming}
\subsubsection{Beaming effect on the emission}
\label{Hydrodynamics-Beaming-emission}
\begin{figure} [h!]
  \centering
  \includegraphics[scale=0.7]{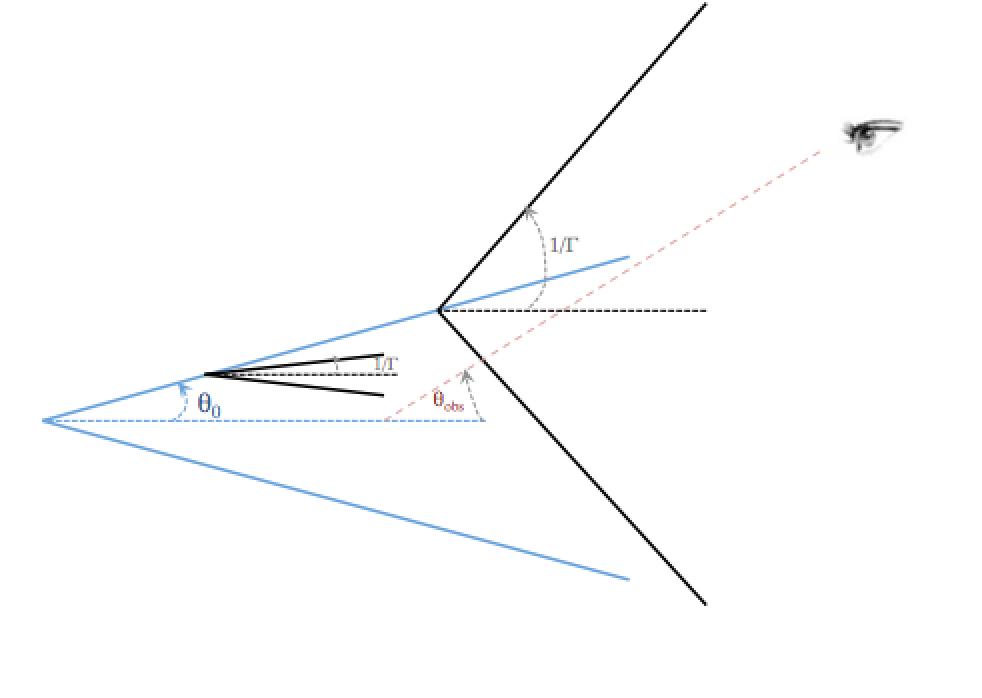}
  \caption{Beaming and observer line of sight effects}
  \label{fig:beaming}
\end{figure}

The above description considers a spherical expansion of the blast wave. Nevertheless, the  radiation from a relativistic source is beamed with a typical beaming angle $1/\gamma $. 
Let $\theta _{0}$ be the half opening angle of the cocoon, see fig. \ \ref{fig:beaming}.
During the relativistic regime, when $\theta _{0}$ is larger than $/1/\gamma$, an observer will see only part of the emission. 
Moreover if the observer is off-jet axis, the observer angle, i.e the angle between jet-axis and line of sight to the observer, $\theta _{obs}$, will affect the radiation received by the observer, see fig . \ \ref{fig:beaming}. 
The following equations present the actual observed flux $F_{obs }$ given the isotropic flux $F_{\nu }$:
\\
\\
If $\frac{1}{\gamma }< \theta _{0}$, we have : 
\begin{equation}
\left\lbrace
                \begin{array}{ll}
                \theta _{obs}<\theta _{0}\, \, \, \, \, \, \, \: \: \: \: \: \: \: F_{obs } = F_{\nu }\\
\theta _{obs}>\theta _{0}\, \, \, \, \, \, \, \: \: \: \: \: \: \: F_{obs } = 0
\end{array}
              \right. .
\end{equation}%

As the Lorentz factor decreases with time, we observe a progressively larger fraction of the emitting region, until ${1}/{\gamma }\approx \theta _{0}$. Then  if ${1}/{\gamma }>\theta _{0}$, we have obtained the following equations to take into account both the beaming effect and the observer position. 
If ${1}/{\gamma }>\theta _{0}$, we have:
\begin{equation}
\left\lbrace
\begin{array}{ll}
\theta_{obs} + \theta _{0} <\frac{1}{\gamma }                   &   F_{obs } = F_{\nu } \\
\theta _{obs} + \theta _{0} >\frac{1}{\gamma }>\theta _{obs}   ~~~~~ &   F_{obs } = F_{\nu } \frac{S\gamma^{2} }{\pi} \\
\frac{1}{\gamma }<\theta _{obs}                                 &   F_{obs } = 0 
\end{array}
\right.
\end{equation}

where $S $ is the surface as shown in fig. 10. Detailed calculation of S are presented in the appendix. 
\begin{figure} [h!]
  \centering
  \includegraphics[scale=0.7]{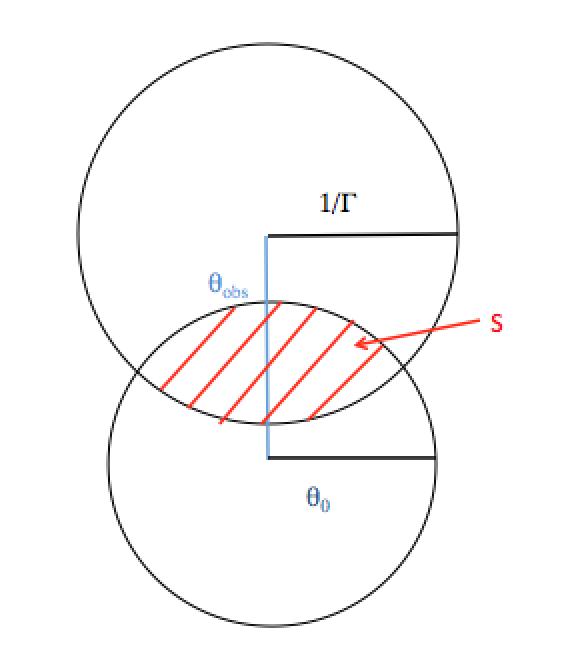}
  \caption{Useful surface of the emission for the observer}
  \label{fig:surface}
\end{figure}

As discussed in the above section, the late time emission of the cocoon afterglow becomes non relativistic  and spherical. Therefore, no relativistic beaming has to be considered, emission seen by the observer only depends on $\theta _{obs}$. 
\newpage
\subsubsection{Beaming effect on the energy}
\label{Hydrodynamics-Beaming-energy}
Beaming also has an effect on luminosity and energy. Indeed, the luminosity $L_{j}$ and  energy $E_{0}$, presented in section \ref{Cocoon-Properties-Energy}, are assuming that the bursts are isotropic. However, when taking beaming into account, the energy $E_{iso}$  has to be considered:
\begin{equation}
E_{iso} = \frac{1}{\theta _{0}^{2} }E_{0}
\end{equation}

Nevertheless, when reaching the non relativistic regime, the expansion is spherical. Hence, the energy that has to be considered is $E_{0}$, as defined in section \ref{Cocoon-Properties-Energy}.

Consequently, in our model we consider $E_{iso}$ for the relativistic regime, and reach $E_{0}$ for the non relativistic one. 

Note that the conical expansion of the half opening angle of the cocoon, $\theta _{0}$, will be taken into account in the complete version of our model.\footnote{Curves with footnotes have been modified to take sideways expansion into account}

\newpage
 
\section{Synchrotron Emission}
\label{Synchrotron}
\subsection{Synchrotron or Inverse Compton}
\label{Synchrotron-Inverse}

Given the fraction of the shock energy going into magnetic energy density, $\varepsilon _{B}$, and into the electron energy density, $\varepsilon_{e}$, as defined in section \ref{Hydrodynamics-Shock-acceleration} , we neglect Compton scattering and consider only synchrotron emission. Indeed, in this model, we consider $\varepsilon_{B}=\varepsilon_{e}=0.1$. (Compton scattering can be important if $\varepsilon_{B}> \varepsilon_{e}$).
\subsection{Discussion about velocity of electrons}
\label{Synchrotron-Discussion}
It is clear that, electrons will be accelerated into higly relativistic Lorentz factor during the Blandford and McKee evolution of the blast wave. Nevertheless, the question of Lorentz factor at which electrons  are accelerated  arises during Sedov and Taylor.
During this phase, the minimum Lorentz factor $\gamma_{m}$ at which electrons are accelerated is given by: 
\begin{equation}
\gamma_{m}=\epsilon_{e}\frac{p-2}{p-1} \frac{m_{p}}{m_{e}}\beta^{2}
\end{equation}

\noindent Where $p =3$.
 
Electrons are not relativistic when $\gamma_{m} \rightarrow  1$ which occurs for $\beta \approx 10^{-2}$. However, we find with our model that while $ t< 10^{3}$ days, we have that $\beta > 10^{-1}$. As a consequence, we can consider that, within reasonable observing time, the electrons are accelerated to relativistic velocities.

\subsection{Synchrotron frequency and power for a relativistic shock}
\label{Synchrotron-Frequency}

Consider a relativistic electron with Lorentz factor $\gamma _{e} \gg 1$ in a magnetic field B, it emits synchrotron radiation. The radiation power and the characteristic frequency for a relativistic shock are given by \citep{Sari1998}:
\begin{equation}
P(\gamma _{e}) = \frac{4}{3} \sigma _{T}c\gamma ^{2}\gamma _{e}^{2}\frac{B^{2}}{8\pi }\beta^{2}
\end{equation}
\begin{equation}
\nu (\gamma _{e}) = \gamma \gamma _{e}^{2}\frac{q_{e}B}{2\pi m_{e}c}
\end{equation}

\noindent where $\sigma _{T}$ is Thomson cross-section , $q_{e}$ the electron charge. 

In the above equations, the factors of $\gamma ^{2}$ and $\gamma $ are used to transform the results from the frame of the shocked fluid to the frame of the observer.

The spectral power, $P_{\nu }$, power per unit frequency, varies as $\nu^{1/3}$ while $ \nu < \nu (\gamma _{e}) $, and cuts off exponentially for $ \nu > \nu (\gamma _{e}) $ \citep{RB}. The peak power occurs at $\nu (\gamma _{e}) $, where it has the approximate value :
\begin{equation}
P_{\nu ,max }\approx\frac{P(\gamma _{e})}{\nu (\gamma _{e})} = \frac{m_{e}c^{2}\sigma _{T}}{3q_{e}}\gamma B
\end{equation}

%faut il un beta^2? pour Pnu, max

\subsection{Synchrotron cooling}
\label{Synchrotron-Cooling}
The above description of $P_{\nu }$ does not take into account the loss of energy due to radiation. However, the electrons  emitting synchrotron radiation are cooling down. The time scale for this to occur is given by the energy of the electrons divided by the rate at which they are radiating away their energy. 

Consider $\gamma _{c}$,  the  critical value above which cooling by synchrotron radiation is significant.  The critical electron Lorentz factor $\gamma _{c}$ is given by the condition \citep{Sari1998}: $$\gamma\gamma _{c} m_{e}c^{2} = P (\gamma _{c})t$$ where $t$ refers to time in the frame of the observer.
\\
\\
Therefore the critical electron Lorentz factor $\gamma _{c}$,in the relativistic regime, is: 
\begin{equation}
\label{eq:sync}
\gamma _{c} = \frac{3 m_{e}}{16 \varepsilon _{B} \sigma _{T}m_{p}c t \gamma ^{3}n_{0}}
\end{equation}

In the non relativistic regime, we obtain a critical electron Lorentz factor $\gamma _{c}$:
\begin{equation}
\gamma _{c} =\frac{3 m_{e} t^{-1} \gamma^{3}}{ 16  \sigma _{T} m_{p} \varepsilon _{B}  n_{0}}
\end{equation}

where $t$ refers to time in the frame of the observer.
In our model, we also compute the value of $\gamma _{c}$ in the transition region.

\subsection{Synchrotron self-absorption}
\label{Synchrotron-Absorption}
Our above calculation assumes that all of the synchrotron radiation emitted by each electron reaches the observer. However this is not necessarily the case: as a photon propagates through the plasma on its way out of the source, there is a chance that it will scatter off one of the synchrotron electrons. This is known as synchrotron self-absorption.
If such scattering occurs many times before the photon can get out of the source, the result is that an outside observer only “sees” emission from a thin layer near the surface of the source. Beneath this, the synchrotron radiation from the electrons are self absorded (i.e the medium is optically thick). For GRB, self absorption may appear at late time and typically in radio emission \citep{Katz1994,Waxman1997}. It leads to a steep cutoff of the low energy spectrum, either as the commonly known $\nu^{5/2}$ or as $\nu^{2}$ . To estimate the self absorption frequency one needs the optical depth along the line of sight. A simple approximation is: $\alpha ^{'}_{\nu ^{'}} R/\gamma$ where $\alpha ^{'}_{\nu ^{'}}$ is the absorption coefficient defined by \cite{Piran2004}: 
\begin{equation}
\alpha ^{'}_{\nu ^{'}} = \frac{p+2}{8\pi m_{e}\nu^{'2} }\int_{\gamma _{min}}^{\infty }d\gamma _{e}P^{'}_{\nu ^{'},e}(\gamma _{e})\frac{n(\gamma _{e})}{\gamma _{e}}
\end{equation}

The self absorption frequency $\nu _{a}$ satisfies: $\alpha ^{'}_{\nu ^{'}} R/\gamma=1$ \citep{Piran2004}.

In the relativistic regime, $\nu _{a}$ is given by \cite{Granot1999}:
\begin{equation}
\nu _{a} = 0.247 \times 4.24 \times 10^{9}\left ( \frac{p+2}{3p+2}\right )^{3/5}\times \frac{(p-1)^{8/5}}{p-2} \varepsilon _{e}^{-1}\varepsilon _{B}^{1/5}E_{52}^{1/5}n_{0}^{3/5}\:  Hz
\end{equation}

\noindent where $E_{52} = E_{0}/10^{52}$.

In the opposite non relativistic regime, where the radio emission will peak,  we consider $\nu _{a}$ \citep{Nakar2011}:
\begin{equation}
\nu _{a}\approx 10^{9} R_{17}^{\frac{2}{p+4}}n_{0}^{\frac{6+p}{2(p+4)}}\varepsilon _{e,-1}^{\frac{2 +p}{2(p+4)}}\varepsilon _{B,-1}^{\frac{2(p-1)}{p+4}}\beta ^{\frac{5p-2}{p+4}}\:  Hz
\end{equation}

In our model, we also compute the value of $\nu _{a}$ in the transition region.
\subsection{Influence of cooling and self-absorption }
\label{Synchrotron-Influence}

As described in section \ref{Hydrodynamics-Shock-acceleration}, electrons are accelerated in the shock to a power law distribution of Lorentz factor $\gamma_e$, with a minimum Lorentz factor $\gamma _{m}$.
In order to observe the impact of both cooling and self-absorption, we need to compare  $\nu_{m} = \nu_{syn}(\gamma _{m})$ with  the cooling frequency $\nu _{c} = \nu_{syn}(\gamma_{c})$ and the self-absorption $\nu_{a}$.
We obtain the following evolution of $\nu _{m}$, $\nu _{c}$ and $\nu _{a}$,  see fig.\ 11.
\begin{figure} [h!]
  \centering
  \includegraphics[scale=0.9]{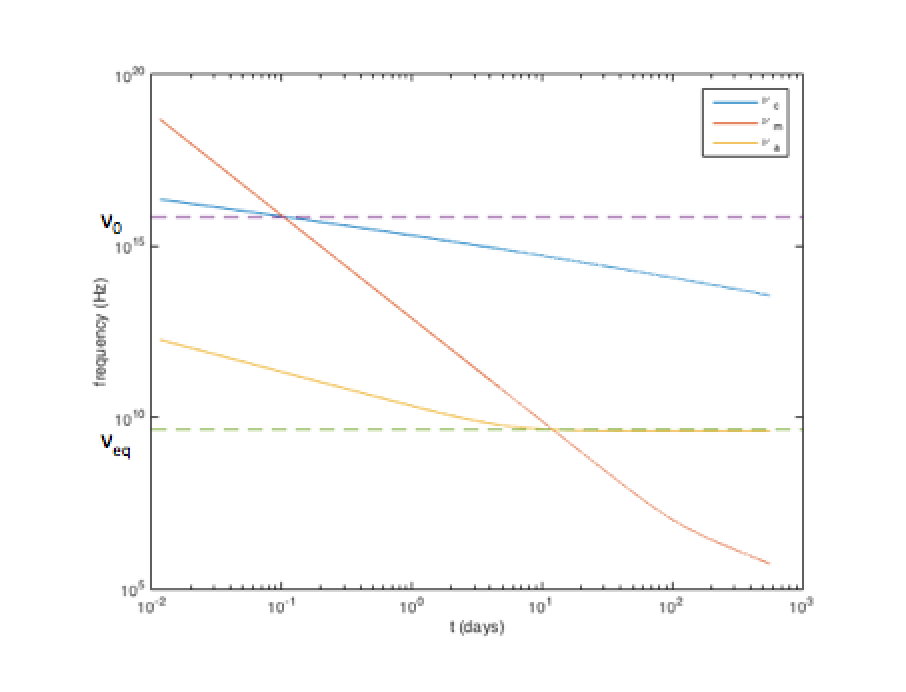}
  \label{figfrequency}
  \caption{Important frequencies for synchrotron emission}

\end{figure}

In the above figure, two important transition frequencies are shown, $\nu_{0}$ and $\nu_{eq}$ where $\nu _{0}$ is the frequency at which $\nu_{m}=\nu_{c}$ and $\nu_{eq}$ for $\nu_{m}=\nu _{a}$.
We can see that $\nu_{0}$ will be important for X-rays emission while $\nu_{eq}$ will play a key role for radio emission.
 
{\color{white}
{asdfasdfa asdfasdfadfs asdfasdfa asdfasdfadfs asdfasdfa asdfasdfadfs asdfasdfa asdfasdfadfs asdfasdfa asdfasdfadfs asdfasdfa asdfasdfadfs asdfasdfa asdfasdfadfs asdfasdfa asdfasdfadfs asdfasdfa asdfasdfadfs asdfasdfa asdfasdfadfs asdfasdfa asdfasdfadfs asdfasdfa asdfasdfadfs asdfasdfa asdfasdfadfs asdfasdfa asdfasdfadfs asdfasdfa asdfasdfadfs asdfasdfa asdfasdfadfs asdfasdfa asdfasdfadfs asdfasdfa asdfasdfadfs asdfasdfa asdfasdfadfs asdfasdfa asdfasdfadfs asdfasdfa asdfasdfadfs asdfasdfa asdfasdfadfs asdfasdfa asdfasdfadfs asdfasdfa asdfasdfadfs asdfasdfa asdfasdfadfs asdfasdfa asdfasdfadfs asdfasdfa asdfasdfadfs asdfasdfa asdfasdfadfs asdfasdfa asdfasdfadfs asdfasdfa asdfasdfadfs asdfasdfa asdfasdfadfs asdfasdfa asdfasdfadfs asdfasdfa asdfasdfadfs asdfasdfa asdfasdfadfs asdfasdfa asdfasdfadfs asdfasdfa asdfasdfadfs asdfasdfa asdfasdfadfs asdfasdfa asdfasdfadfs asdfasdfa asdfasdfadfs asdfasdfa asdfasdfadfs asdfasdfa asdfasdfadfs asdfasdfa asdfasdfadfs asdfasdfa asdfasdfadfs asdfasdfa asdfasdfadfs asdfasdfa asdfasdfadfs asdfasdfa asdfasdfadfs asdfasdfa asdfasdfadfs asdfasdfa asdfasdfadfs asdfasdfa asdfasdfadfs}

{asdfasdfa asdfasdfadfs asdfasdfa asdfasdfadfs asdfasdfa asdfasdfadfs asdfasdfa asdfasdfadfs asdfasdfa asdfasdfadfs asdfasdfa asdfasdfadfs asdfasdfa asdfasdfadfs asdfasdfa asdfasdfadfs asdfasdfa asdfasdfadfs asdfasdfa asdfasdfadfs asdfasdfa asdfasdfadfs asdfasdfa asdfasdfadfs asdfasdfa asdfasdfadfs asdfasdfa asdfasdfadfs asdfasdfa asdfasdfadfs asdfasdfa asdfasdfadfs asdfasdfa asdfasdfadfs asdfasdfa asdfasdfadfs asdfasdfa asdfasdfadfs asdfasdfa asdfasdfadfs asdfasdfa asdfasdfadfs asdfasdfa asdfasdfadfs asdfasdfa asdfasdfadfs asdfasdfa asdfasdfadfs asdfasdfa asdfasdfadfs asdfasdfa asdfasdfadfs asdfasdfa asdfasdfadfs asdfasdfa asdfasdfadfs asdfasdfa asdfasdfadfs asdfasdfa asdfasdfadfs asdfasdfa asdfasdfadfs asdfasdfa asdfasdfadfs asdfasdfa asdfasdfadfs asdfasdfa asdfasdfadfs asdfasdfa asdfasdfadfs asdfasdfa asdfasdfadfs asdfasdfa asdfasdfadfs asdfasdfa asdfasdfadfs asdfasdfa asdfasdfadfs asdfasdfa asdfasdfadfs asdfasdfa asdfasdfadfs asdfasdfa asdfasdfadfs asdfasdfa asdfasdfadfs asdfasdfa asdfasdfadfs asdfasdfa asdfasdfadfs asdfasdfa asdfasdfadfs asdfasdfa asdfasdfadfs asdfasdfa asdfasdfadfs asdfasdfa asdfasdfadfs}}

\section{Light curves}
 \label{Light}
\subsection{Spectrum for X-ray and Optical emission}
 \label{Light-Xray}
\subsubsection{Fast cooling } 
 \label{Light-Xray-Fast}
As described in section \ref{Hydrodynamics-Shock-acceleration}, electrons are accelerated in the shock to a power law distribution of Lorentz factor $\gamma_e$, with a minimum Lorentz factor $\gamma _{m} $.
To calculate the net spectrum due to all the electrons we need to integrate over $\gamma _{e}$. Let the total number of electrons accelerated be $N_{e}$.
 If $\gamma _{m}  > \gamma _{c} $, all the electrons cool down roughly to $\gamma _{c} $ and the flux at $\nu _{c} $ is approximately $N_{e} P_{\nu,max }$ .  We call this the case of fast cooling. The isotropic flux at the observer, $F_{\nu }$, is given by \cite{Sari1998}: 
\begin{equation}
 F_{\nu } =\left\{
                \begin{array}{ll}
                 \left (  \frac{\nu }{\nu _{c}} \right )^{1/3}F_{\nu, max }\: \: \: \: \: \: \: \: \:  \: \: \: \:  \: \: \nu _{c}>\nu\\
\left (  \frac{\nu }{\nu _{c}} \right )^{-1/2}F_{\nu, max }\: \: \: \: \: \: \: \: \:  \: \: \: \:  \: \: \nu _{m}>\nu>\nu _{c}\\
\left (  \frac{\nu_{m} }{\nu _{c}} \right )^{-1/2}\left (  \frac{\nu }{\nu _{m}} \right )^{-p/2}F_{\nu, max }\: \: \: \: \: \: \: \: \:  \: \: \: \:  \: \: \nu>\nu _{m}
\end{array}
              \right.
 \end{equation} 
 
\noindent where $F_{\nu, max }= N_{e} P_{\nu,max }/4 \pi D^{2}$ is the observed peak flux at distance D from the source considering an isotropic flux.
\subsubsection{Slow cooling } 
 \label{Light-Xray-Slow}
When $\gamma _{c}  > \gamma _{m} $, only the electrons with $\gamma _{e}  > \gamma _{c} $. We call this slow cooling, because the electrons with $\gamma _{e}  \approx \gamma _{c} $, which represent a large fraction of the electron population, do not cool within a time $t$, see eq.\ \ref{eq:sync}. Integrating over the electron distribution, we have the following isotropic flux at the observer, $F_{\nu }$, is given by \cite{Sari1998}: 
\begin{equation}
F_{\nu } =\left\{
                \begin{array}{ll}
                 \left (  \frac{\nu }{\nu _{m}} \right )^{1/3}F_{\nu, max }\: \: \: \: \: \: \: \: \:  \: \: \: \:  \: \: \nu _{m}>\nu\\
\left (  \frac{\nu }{\nu _{m}} \right )^{-(p-1)/2}F_{\nu, max }\: \: \: \: \: \: \: \: \:  \: \: \: \:  \: \: \nu _{c}>\nu>\nu _{m}\\
\left (  \frac{\nu_{c} }{\nu _{m}} \right )^{-(p-1)/2}\left (  \frac{\nu }{\nu _{c}} \right )^{-p/2}F_{\nu, max }\: \: \: \: \: \: \: \: \:  \: \: \: \:  \: \: \nu>\nu _{c}
\end{array}
              \right.
\end{equation}

\subsection{Light curves for X-ray and Optical emission}
 \label{Light curves for X-ray and Optical emission}
The instantaneous spectra described in the previous section do not depend on the hydrodynamical evolution of the shock. Nevertheless, the light curves at a given frequency, do  depend on hydrodynamics evolution. As shown, in section, $ N_{e} $ and $\gamma$ vary with time. We have also shown that $\nu _{m}$ and $\nu_{c}$ vary with time in section \ref{Synchrotron-Influence} .
Our model which takes into account the full hydrodynamics evolution in BM , transition and ST regime allows us to calculate the light curves.
Moreover as shown in section \ref{Hydrodynamics-Beaming}, with our model the beaming effect are taken into account to compute the flux seen by the observer. 
The beaming also affects the number of accelerated electrons $N_{e}$. Indeed, during the relativistic phase where there is beaming  $N_{e} = n_{0}\pi \theta_{0}^{2}R^{3}$ while during the non relativistic phase where there is no beaming $N_{e} = n_{0}\frac{4\pi }{3}R^{3}$.

Finally, given an observer at a distance $D = 10 ^{28}$ cm , with an half opening angle of the cocoon $\theta_0 = 20 \text{deg}$, a density $n=1 cm^{-3} $, an energy $ E = 10 ^50 $erg at the given frequency $\nu = 7. \:10^{14}$ Hz and $\nu = 6.\: 10^{16}$ we obtain the following light curves for both on-axis and off-axis observer :

\begin{figure}[p]
  \centering
  \includegraphics[scale=0.65]{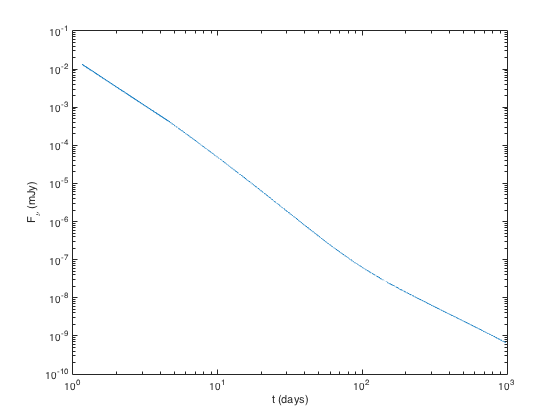}
  \caption[bla]{Optical light curve on-axis observer \footnotemark}
\end{figure}
\footnotetext{\label{foot_curve} curve modified after Master's submission to integrate sideways expansion}
\begin{figure}[p]
  \centering
  \includegraphics[scale=0.65]{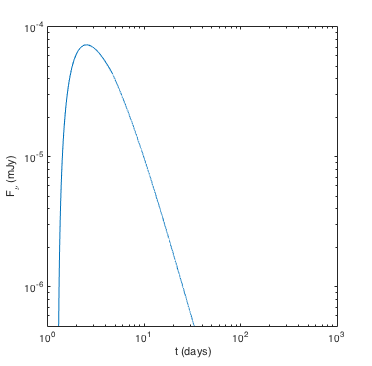}
  \caption[bla]{Optical light curve off-axis observer \footnoteref{foot_curve} }
\end{figure}

\begin{figure}[p]
  \centering
  \includegraphics[scale=0.65]{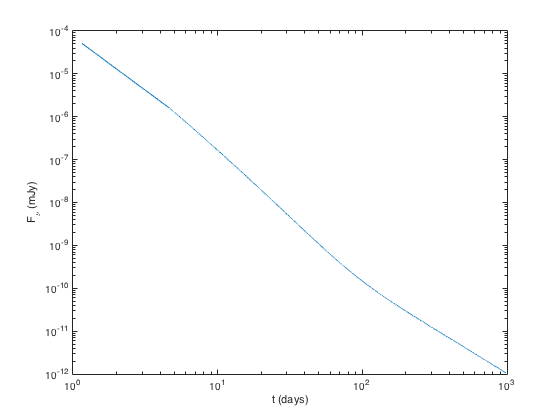}
  \caption[bla]{X-ray light curve on-axis observer \footnoteref{foot_curve}}
\end{figure}
\begin{figure}[p]
  \centering
  \includegraphics[scale=0.65]{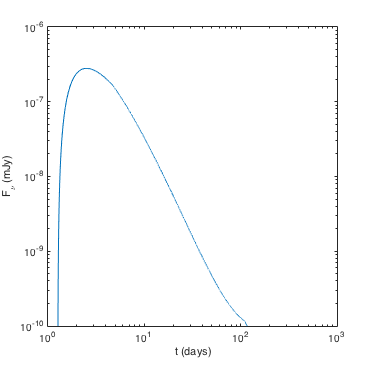}
  \caption[bla]{X-ray light curve off-axis observer \footnoteref{foot_curve}}
\end{figure}

For both X-ray and optical emission an off-axis observer begins to "see" the flux later because of her position and the flux observed during the relativistic beamed part is less important than previously because of the beaming effects discussed in section \ref{Hydrodynamics-Beaming}.

{\color{white}
{asdfasdfa asdfasdfadfs asdfasdfa asdfasdfadfs asdfasdfa asdfasdfadfs asdfasdfa asdfasdfadfs asdfasdfa asdfasdfadfs asdfasdfa asdfasdfadfs asdfasdfa asdfasdfadfs asdfasdfa asdfasdfadfs asdfasdfa asdfasdfadfs asdfasdfa asdfasdfadfs asdfasdfa asdfasdfadfs asdfasdfa asdfasdfadfs asdfasdfa asdfasdfadfs asdfasdfa asdfasdfadfs asdfasdfa asdfasdfadfs asdfasdfa asdfasdfadfs asdfasdfa asdfasdfadfs asdfasdfa asdfasdfadfs asdfasdfa asdfasdfadfs asdfasdfa asdfasdfadfs asdfasdfa asdfasdfadfs asdfasdfa asdfasdfadfs asdfasdfa asdfasdfadfs asdfasdfa asdfasdfadfs asdfasdfa asdfasdfadfs asdfasdfa asdfasdfadfs asdfasdfa asdfasdfadfs asdfasdfa asdfasdfadfs asdfasdfa asdfasdfadfs asdfasdfa asdfasdfadfs asdfasdfa asdfasdfadfs asdfasdfa asdfasdfadfs asdfasdfa asdfasdfadfs asdfasdfa asdfasdfadfs asdfasdfa asdfasdfadfs asdfasdfa asdfasdfadfs asdfasdfa asdfasdfadfs asdfasdfa asdfasdfadfs asdfasdfa asdfasdfadfs asdfasdfa asdfasdfadfs asdfasdfa asdfasdfadfs asdfasdfa asdfasdfadfs asdfasdfa asdfasdfadfs asdfasdfa asdfasdfadfs asdfasdfa asdfasdfadfs asdfasdfa asdfasdfadfs asdfasdfa asdfasdfadfs asdfasdfa asdfasdfadfs asdfasdfa asdfasdfadfs}}

{\color{white}
{asdfasdfa asdfasdfadfs asdfasdfa asdfasdfadfs asdfasdfa asdfasdfadfs asdfasdfa asdfasdfadfs asdfasdfa asdfasdfadfs asdfasdfa asdfasdfadfs asdfasdfa asdfasdfadfs asdfasdfa asdfasdfadfs asdfasdfa asdfasdfadfs asdfasdfa asdfasdfadfs asdfasdfa asdfasdfadfs asdfasdfa asdfasdfadfs asdfasdfa asdfasdfadfs asdfasdfa asdfasdfadfs asdfasdfa asdfasdfadfs asdfasdfa asdfasdfadfs asdfasdfa asdfasdfadfs asdfasdfa asdfasdfadfs asdfasdfa asdfasdfadfs asdfasdfa asdfasdfadfs asdfasdfa asdfasdfadfs asdfasdfa asdfasdfadfs asdfasdfa asdfasdfadfs asdfasdfa asdfasdfadfs asdfasdfa asdfasdfadfs asdfasdfa asdfasdfadfs asdfasdfa asdfasdfadfs asdfasdfa asdfasdfadfs asdfasdfa asdfasdfadfs asdfasdfa asdfasdfadfs asdfasdfa asdfasdfadfs asdfasdfa asdfasdfadfs asdfasdfa asdfasdfadfs asdfasdfa asdfasdfadfs asdfasdfa asdfasdfadfs asdfasdfa asdfasdfadfs asdfasdfa asdfasdfadfs asdfasdfa asdfasdfadfs asdfasdfa asdfasdfadfs asdfasdfa asdfasdfadfs asdfasdfa asdfasdfadfs asdfasdfa asdfasdfadfs asdfasdfa asdfasdfadfs asdfasdfa asdfasdfadfs asdfasdfa asdfasdfadfs asdfasdfa asdfasdfadfs asdfasdfa asdfasdfadfs asdfasdfa asdfasdfadfs asdfasdfa asdfasdfadfs}}

%\newpage

\subsection{Spectrum for radio }
 \label{Light-Radio-Spectrum}
 As discussed in section \ref{Synchrotron-Influence}, for radio spectrum, the two important frequencies are $\nu _{m}$ and $\nu _{a}$. Similarly than for X-ray spectrum, there are two distincts cases : $\nu _{m}>\nu _{a}$ and $\nu _{m}<\nu _{a}$. Integrating over the electron distribution, we have the following isotropic flux at the observer, $F_{\nu }$, \citep{Nakar2011}: 
 
If $\nu _{m}>\nu _{a}$,
\begin{equation}
F_{\nu } =\left\{
                \begin{array}{ll}
                 \left (  \frac{\nu }{\nu _{a}} \right )^{2}\left (  \frac{\nu_{a} }{\nu _{m}} \right )^{1/3}F_{m }\: \: \: \: \: \: \: \: \:  \: \: \: \:  \: \: \nu _{a}>\nu\\
\left (  \frac{\nu }{\nu _{m}} \right )^{1/3}F_{m }\: \: \: \: \: \: \: \: \:  \: \: \: \:  \: \: \nu _{m}>\nu>\nu _{a}\\
\left (  \frac{\nu }{\nu _{m}} \right )^{-(p-1)/2}F_{m }\: \: \: \: \: \: \: \: \:  \: \: \: \:  \: \: \nu>\nu _{m}
\end{array}
              \right.
\end{equation}

\noindent with $F_{ m}\:\approx\:  0.5 \: mJy \:\:R_{17}^{3 }\: n_{0}^{3/2 } \:\varepsilon _{B}^{1/2 }\:\beta \:D_{27}^{-2}$ where $R_{17} = R/10^{17}\:$ cm and $D_{27} = D/10^{27}\:$ cm. 

If $\nu _{m}>\nu _{a}$, $F_{m}$ is non real i.e.  the spectrum does not peak at $F_{m }$ but it peaks at  $F_{a}\:\approx\: \left (  \frac{\nu _{a}}{\nu _{m}} \right )^{-(p-1)/2}F_{m }   $ .
Integrating over the electron distribution, we have the following isotropic flux at the observer, $F_{\nu }$,\citep{Nakar2011}:

\begin{equation}
F_{\nu } =\left\{
                \begin{array}{ll}
                 \left (  \frac{\nu }{\nu _{m}} \right )^{-(p-1)/2}F_{m }\: \: \: \: \: \: \: \: \:  \: \: \: \:  \: \: \nu >\nu_{a}\\
\left (  \frac{\nu }{\nu _{a}} \right )^{5/2}F_{a }\: \: \: \: \: \: \: \: \:  \: \: \: \:  \: \: \nu _{a}>\nu>\nu _{m}\\
\left (  \frac{\nu_{m} }{\nu _{a}} \right )^{5/2}\left (  \frac{\nu }{\nu _{m}} \right )^{2}F_{a }\: \: \: \: \: \: \: \: \:  \: \: \: \:  \: \: \nu<\nu _{m}
\end{array}
              \right.
\end{equation}

\subsection{Light curves for radio emission}
 \label{Light-Radio-Light}
Similarly to section \ref{Light curves for X-ray and Optical emission}, given an observer at a distance $D = 10 ^{28}$ cm , a density $n=1 cm^{-3} $, an energy $ E = 10 ^50 $erg a frequency $\nu = 1\:10^{10}$ Hz we obtain the following light curves for both on-axis and off-axis observer. The  opening angle of the cocoon is $\theta_0 = 20 \:\text{deg}$ and the observer angle is $\theta_0 = 40\: \text{deg}$.
\begin{figure}[h!]
  \centering
  \includegraphics[scale=0.7]{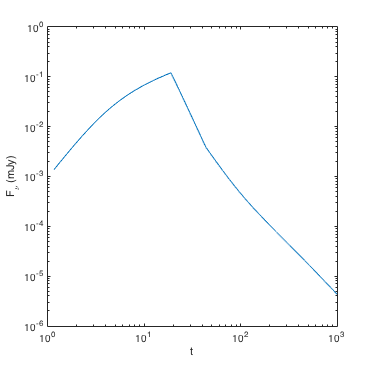}
  \caption[bla]{Radio light curve on-axis observer \footnoteref{foot_curve}}
  \label{fig::LCradio}
\end{figure}
\begin{figure}[h!]
  \centering
  \includegraphics[scale=0.7]{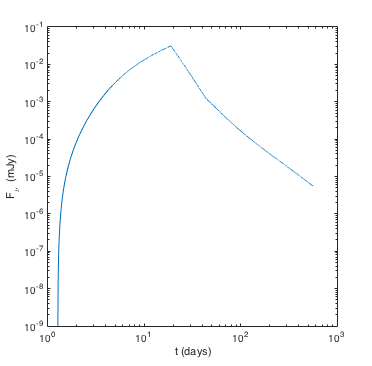}
  \caption[bla]{Radio light curve off-axis observer \footnoteref{foot_curve}}
  \label{fig::LCradiobeaming}
\end{figure}

Similarly to section \ref{Light curves for X-ray and Optical emission}, an off-axis observer begins to "see" emission later and the flux observed for the relativistic part is less important.
It can be seen on both curves that the peak occurs at 20 days. At this time, $\nu_m$ and $\nu_a$  are equal so the spectrum calculation changed as discussed in section \ref{Light-Radio-Spectrum}. The later change in the slope occurs at the time when $\nu_m$ is equal to $\nu$  which also induce a change in the sepctrum as discussed in section \ref{Light-Radio-Spectrum}.

\newpage
\section{Is the cocoon afterglow a promising EM counterpart of GW ?}
 \label{Is the cocoon afterglow a promising EM counterpart of GW ?}
Hereafter we firstly present the binary neutron star merger detectability, then discuss the other EM counterpart candidates. Finally, we compare the cocoon afterglow to the GRB afterglow.
\subsection{Neutron star binary merger rate}
 \label{Can LIGO observe neutron star binary merger?}
To date, the three events detected by LIGO are binary black hole mergers \citep{Abbott2017,GW1,GW2}. Except perhaps in rare circumstances, the merger of stellar mass black holes are not expected to produce luminous EM emission due to the absence of baryonic matter in these systems. Nevertheless, as shown in the following figure fig. \ref{fig::NSNS} , GW emitted during binary neutron star mergers will be above the sensitivity threshold of Advanced LIGO for the next campaigns. Therefore looking for EM counterpart of binary neutron star merger will hopefully allow localization of future GW emitted during such events. 
\begin{figure}[h!]
  \centering
  \includegraphics[scale=0.7]{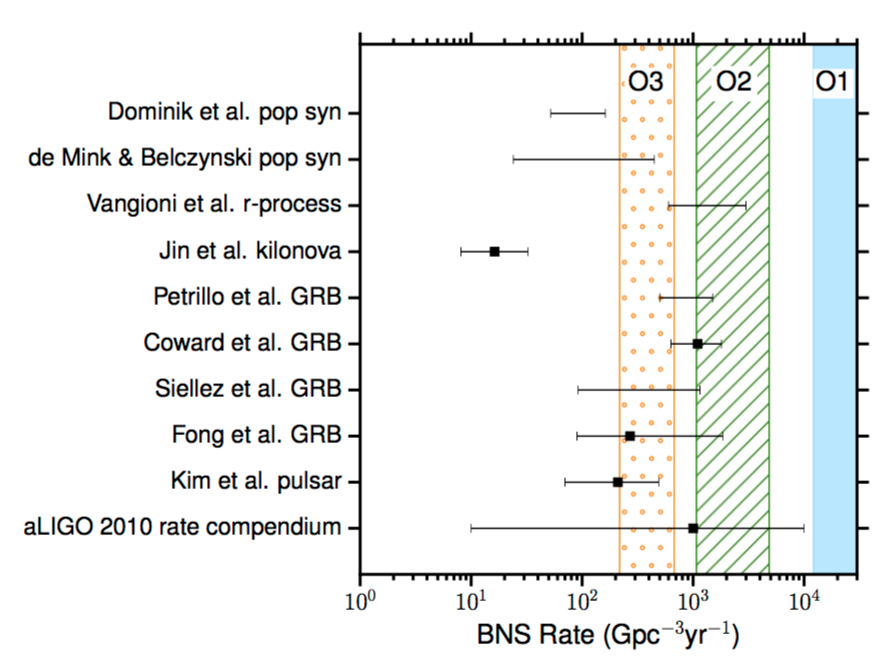}
  \caption{Binary neutron star merger rate threshold for current and futur LIGO campaigns}
  \label{fig::NSNS}
\end{figure}
\newpage
\subsection{Discussion about other possible EM counterparts }
 \label{IsTheCocoon-Discussion}
Macronova, r-process supernova like event and radio flares discussed in section  \ref{Intro-Motivation}are other EM candidates. They also arise from binary neutron star merger.

Matter ejected during binary neutron star merger is enriched by heavy unstable nuclei whose radioactive decay power a macronova  \citep{Li1998,Kulkarni2005,Metzger2010}. However, the macronova will peak in infrared, which make them less likely to be observed considering infrared  telescope sensitivities.
 
The interaction of the expanding ejecta, produced in binary neutron star merger, with the surrounding medium produces, at a later stage, a radio flare lasting months to years, but peaking around a year after the prompt emission \citep{Nakar2011}.
\subsection{Comparison with GRB afterglow}
 \label{IsTheCocoon-Comparison}
The cocoon afterglow is produced by the same physical mechanisms than the GRB afterglow. Consequently, our model can be used to calculate light curves for the GRB afterglow. For the GRB afterglow, we consider the same energy than for the cocoon afterglow, see section  \ref{Cocoon-Properties-Energy}, an initial Lorentz factor $\gamma_{0} \approx 200 $  and a half opening of the jet $\theta _{0}$ = 10 deg.
 \subsubsection{X-ray emission}
For a frequency of $\nu = 6. \: 10^{16}$ Hz, we obtain the following light curves for the cocoon afterglow and the GRB afterglow see fig.\ 19 and fig. \ 20.
\begin{figure}[p]
  \centering
  \includegraphics[scale=0.7]{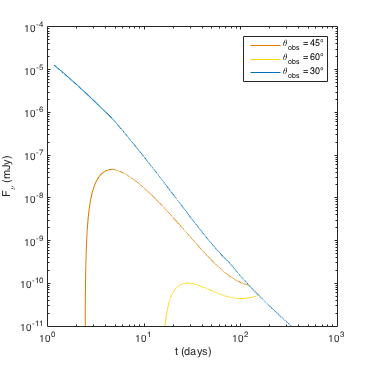}
  \caption[bla]{X-ray emission of the cocoon afterglow \footnoteref{foot_curve}}
  \label{fig::6-xraycocoon}
\end{figure}

\begin{figure}[p]
\centering
  \includegraphics[scale=0.7]{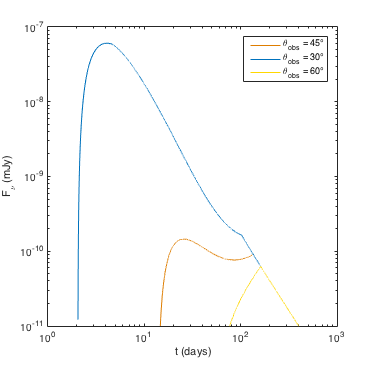}
   \caption[bla]{X-ray emission of the GRB afterglow \footnoteref{foot_curve}}
  \label{fig::6-xrayjet}
  \end{figure}

It can be observed that the cocoon afterglow is both brighter and appears sooner than the orphan jet afterglow while being off-axis. For both afterglows, a more important observer angle gives a later and less bright emission.

 \subsubsection{Optical emission}
  \label{IsTheCocoon-Comparison-Optical}
For a frequency of $\nu = 7. \: 10^{14}$ Hz, we obtain the following light curves for the cocoon afterglow and the GRB afterglow see fig.\ 21 and fig.\ 22.
\begin{figure}[p]
  \centering
  \includegraphics[scale=0.7]{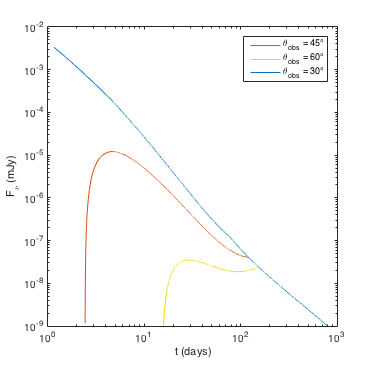}
   \caption[bla]{Optical emission of the cocoon afterglow \footnoteref{foot_curve}}  
  \label{fig::optical_cocoon_6}
\end{figure}

\begin{figure}[p]
\centering
  \includegraphics[scale=0.7]{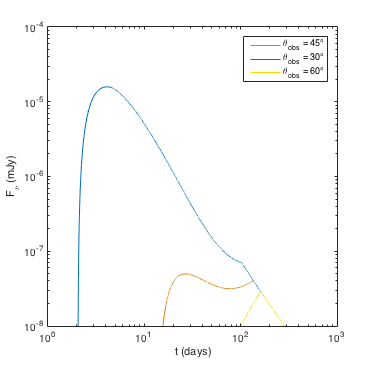}
   \caption[bla]{Optical emission of the GRB afterglow \footnoteref{foot_curve}}
  \label{fig::6opticaljet}
  \end{figure}
  
Similarly to X-ray emission, it can be observed that the cocoon afterglow is both brighter and appears sooner than the orphan jet afterglow while being off-axis. For both afterglows, a more important the observer angle gives a later and the less bright emission.

\subsubsection{Radio emission}
  \label{IsTheCocoon-Comparison-Radio}
For a frequency of $\nu = 1 $GHz, we obtain the following light curves for the cocoon afterglow and the GRB afterglow see fig.\ 23 and fig.\ 24.
\begin{figure}[p]
  \centering
  \includegraphics[scale=0.7]{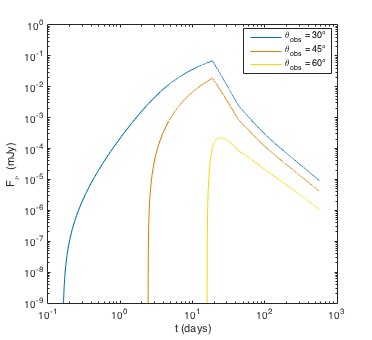}
   \caption[bla]{Radio emission of the cocoon afterglow \footnoteref{foot_curve}}
   \label{fig::6-10GHz-cocoon}
\end{figure}

\begin{figure}[p]
\centering
  \includegraphics[scale=0.7]{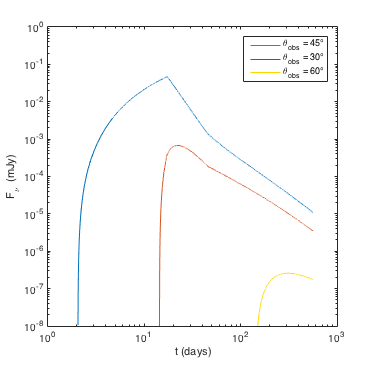}
   \caption[bla]{Radio emission of the GRB afterglow \footnoteref{foot_curve}}
  \label{fig::6-10GHzjet}
  \end{figure}
  
Differently than before, for an observer angle of 30 deg, both afterglow emissions are comparable. However, the cocoon afterlow occurs one day before the jet afterglow.
For an observer angle of 45 deg or 60 deg, the orphan jet afterglow is less bright than the cocoon one and appears later.

\newpage
\section{Conclusion}
\label{Conclusion}
The detection of a GW event with a coincidental EM counterpart will allow the localization of the event progenitor. This should provide us with important information about one of the most intriguing and energetic phenomena in our Universe, that of neutron star mergers. It will allow us to ascertain the effective sensitivity of GW detectors.

The EM candidates : macronova and radio flares,  discussed  in section \ref{IsTheCocoon-Comparison} exhibit several uncertain characteristics making their possible observation in doubt. Therefore, we considered in this work another a less known and non studied EM counterpart : the cocoon afterglow. For that purpose, we propose a model that provides the full hydrodynamic evolution of a blast wave including the mildly relativistic regime where neither Sedov-Taylor solution nor its fully relativistic counterpart Blandford-McKee are valid.

As shown, in section \ref{IsTheCocoon-Comparison}, under favorable conditions the cocoon afterglow emission is comparable to the GRB afterglow. However unlike the latter, it will be observable, depending on the observer angle, a few days to more than a dozen of days before the orphan GRB afterglow itself. \textbf {Therefore, we expect the signal arising from the cocoon afterglow to be of prime importance}.

In our subsequent study, we will consider more sophisticated hypothesis for sideways emission.
 %the conical evolution for the half opening angle of the cocoon \citep{Granot2012}. 
 %We will also consider a broader range of hypotheses for both the Lorentz factors and energies.  
 We can conclude that the cocoon afterglow can be a promising EM counterpart and will be our future research project.

\newpage
\section{Appendix: Beaming issue of the surface S}
\begin{equation}
\begin{split}
\theta_{bs} &= \frac{1}{\Gamma} + \theta_j - d \\
\Longrightarrow	d &= \frac{1}{\Gamma} + \theta_j - \theta_{bs}
\end{split}
\end{equation}
Pythagore:
\begin{equation}
\begin{split}
\frac{1}{r^2} = \left( \frac{1}{\Gamma} - \frac{d}{2}  \right)^2 + d_2^2
\Longrightarrow	d_2 &=\left( \frac{d}{\Gamma} - \frac{d^2}{4} \right)^{\frac{1}{2}}
\end{split}
\end{equation}
($d_2$ full diagonal)
\begin{equation}
A = d \times \frac{d_2}{2}
\end{equation}
\begin{equation}
A = \frac{d}{2} \times \left( \frac{d}{\Gamma} - \frac{d^2}{4} \right)^{\frac{1}{2}}
\end{equation}
\def\ref@jnl#1{{#1}}

\def\aj{\ref@jnl{AJ}}                   % Astronomical Journal
\def\actaa{\ref@jnl{Acta Astron.}}      % Acta Astronomica
\def\araa{\ref@jnl{ARA\&A}}             % Annual Review of Astron and Astrophys
\def\apj{\ref@jnl{ApJ}}                 % Astrophysical Journal
\def\apjl{\ref@jnl{ApJ}}                % Astrophysical Journal, Letters
\def\apjs{\ref@jnl{ApJS}}               % Astrophysical Journal, Supplement
\def\ao{\ref@jnl{Appl.~Opt.}}           % Applied Optics
\def\apss{\ref@jnl{Ap\&SS}}             % Astrophysics and Space Science
\def\aap{\ref@jnl{A\&A}}                % Astronomy and Astrophysics
\def\aapr{\ref@jnl{A\&A~Rev.}}          % Astronomy and Astrophysics Reviews
\def\aaps{\ref@jnl{A\&AS}}              % Astronomy and Astrophysics, Supplement
\def\azh{\ref@jnl{AZh}}                 % Astronomicheskii Zhurnal
\def\baas{\ref@jnl{BAAS}}               % Bulletin of the AAS
\def\bac{\ref@jnl{Bull. astr. Inst. Czechosl.}}
                % Bulletin of the Astronomical Institutes of Czechoslovakia 
\def\caa{\ref@jnl{Chinese Astron. Astrophys.}}
                % Chinese Astronomy and Astrophysics
\def\cjaa{\ref@jnl{Chinese J. Astron. Astrophys.}}
                % Chinese Journal of Astronomy and Astrophysics
\def\icarus{\ref@jnl{Icarus}}           % Icarus
\def\jcap{\ref@jnl{J. Cosmology Astropart. Phys.}}
                % Journal of Cosmology and Astroparticle Physics
\def\jrasc{\ref@jnl{JRASC}}             % Journal of the RAS of Canada
\def\memras{\ref@jnl{MmRAS}}            % Memoirs of the RAS
\def\mnras{\ref@jnl{MNRAS}}             % Monthly Notices of the RAS
\def\na{\ref@jnl{New A}}                % New Astronomy
\def\nar{\ref@jnl{New A Rev.}}          % New Astronomy Review
\def\pra{\ref@jnl{Phys.~Rev.~A}}        % Physical Review A: General Physics
\def\prb{\ref@jnl{Phys.~Rev.~B}}        % Physical Review B: Solid State
\def\prc{\ref@jnl{Phys.~Rev.~C}}        % Physical Review C
\def\prd{\ref@jnl{Phys.~Rev.~D}}        % Physical Review D
\def\pre{\ref@jnl{Phys.~Rev.~E}}        % Physical Review E
\def\prl{\ref@jnl{Phys.~Rev.~Lett.}}    % Physical Review Letters
\def\pasa{\ref@jnl{PASA}}               % Publications of the Astron. Soc. of Australia
\def\pasp{\ref@jnl{PASP}}               % Publications of the ASP
\def\pasj{\ref@jnl{PASJ}}               % Publications of the ASJ
\def\rmxaa{\ref@jnl{Rev. Mexicana Astron. Astrofis.}}%
                % Revista Mexicana de Astronomia y Astrofisica
\def\qjras{\ref@jnl{QJRAS}}             % Quarterly Journal of the RAS
\def\skytel{\ref@jnl{S\&T}}             % Sky and Telescope
\def\solphys{\ref@jnl{Sol.~Phys.}}      % Solar Physics
\def\sovast{\ref@jnl{Soviet~Ast.}}      % Soviet Astronomy
\def\ssr{\ref@jnl{Space~Sci.~Rev.}}     % Space Science Reviews
\def\zap{\ref@jnl{ZAp}}                 % Zeitschrift fuer Astrophysik
\def\nat{\ref@jnl{Nature}}              % Nature
\def\iaucirc{\ref@jnl{IAU~Circ.}}       % IAU Cirulars
\def\aplett{\ref@jnl{Astrophys.~Lett.}} % Astrophysics Letters
\def\apspr{\ref@jnl{Astrophys.~Space~Phys.~Res.}}
                % Astrophysics Space Physics Research
\def\bain{\ref@jnl{Bull.~Astron.~Inst.~Netherlands}} 
                % Bulletin Astronomical Institute of the Netherlands
\def\fcp{\ref@jnl{Fund.~Cosmic~Phys.}}  % Fundamental Cosmic Physics
\def\gca{\ref@jnl{Geochim.~Cosmochim.~Acta}}   % Geochimica Cosmochimica Acta
\def\grl{\ref@jnl{Geophys.~Res.~Lett.}} % Geophysics Research Letters
\def\jcp{\ref@jnl{J.~Chem.~Phys.}}      % Journal of Chemical Physics
\def\jgr{\ref@jnl{J.~Geophys.~Res.}}    % Journal of Geophysics Research
\def\jqsrt{\ref@jnl{J.~Quant.~Spec.~Radiat.~Transf.}}
                % Journal of Quantitiative Spectroscopy and Radiative Transfer
\def\memsai{\ref@jnl{Mem.~Soc.~Astron.~Italiana}}
                % Mem. Societa Astronomica Italiana
\def\nphysa{\ref@jnl{Nucl.~Phys.~A}}   % Nuclear Physics A
\def\physrep{\ref@jnl{Phys.~Rep.}}   % Physics Reports
\def\physscr{\ref@jnl{Phys.~Scr}}   % Physica Scripta
\def\planss{\ref@jnl{Planet.~Space~Sci.}}   % Planetary Space Science
\def\procspie{\ref@jnl{Proc.~SPIE}}   % Proceedings of the SPIE

\let\astap=\aap
\let\apjlett=\apjl
\let\apjsupp=\apjs
\let\applopt=\ao

\newpage
\bibliographystyle{authordate1}                
\bibliography{bibliographie}{}

\end{document}